# 1.5 Million Messages Per Second on 3 Machines: Benchmarking and Latency Optimization of Apache Pulsar at Enterprise Scale

Muhamed Ramees Cheriya Mukkolakkal

*Platform Engineering, Brivo Systems*

*Austin, Texas, United States*

rameescm1@gmail.com

**Abstract**—We present a two-phase investigation into Apache Pulsar latency optimization and large-scale throughput validation on Kubernetes bare-metal infrastructure. Three independent root causes were identified via Java Flight Recorder (JFR) profiling: (1) G1GC pauses with 32 GB heap, (2) journal fdatasync on worn SSDs averaging 5.1 ms, and (3) a previously undocumented Linux kernel page cache writeback interaction inside BookKeeper's ForceWriteThread that degrades journal fdatasync from <1 ms to 15–22 ms even across physically separate NVMe drives sharing the kernel block layer. We validated **1,499,947 msg/s at 3.88 ms median publish latency with zero failures** on 3 bare-metal nodes over 10 minutes. We project 15 M msg/s on 15 machines via Pulsar's native key-based partition routing with no external load balancer.

## I. INTRODUCTION

Apache Pulsar [1] is a cloud-native distributed messaging platform built on Apache BookKeeper [2] for durable tiered storage. Unlike Kafka, Pulsar separates serving (brokers) from storage (bookies), enabling independent horizontal scaling. Brivo Systems operates multi-cluster Pulsar deployments for real-time event streaming across physical security infrastructure, processing millions of device events daily on Kubernetes.

Despite running at moderate loads (700–9,000 msg/s), production clusters exhibited 13–18 ms median publish latency and intermittent 213 ms spikes, motivating a systematic investigation combining JVM profiling, OS-level analysis, and controlled benchmarking. This paper makes three primary contributions:

- Identification of three independent latency root causes via JFR profiling on live bookie nodes, including a previously undocumented kernel writeback interaction in BookKeeper's ForceWriteThread.
- A validated benchmark achieving **1.5 M msg/s on 3 bare-metal nodes** at 3.88 ms median publish latency with zero failures over 10 minutes, with 65–82% headroom on all non-network resources.
- A projected architecture reaching **15 M msg/s on 15 machines** (~1.3 trillion messages/day) using Pulsar's native key-based partition routing with no external load balancer and full per-cluster fault isolation.

Section II presents production cluster analysis. Section III details root cause analysis. Section IV describes optimization experiments. Section V presents the benchmark. Section VI covers scaling architecture. Sections VII and VIII present related work and conclusions.

## II. PRODUCTION CLUSTER ANALYSIS

### A. Cluster Overview

Four production Pulsar 4.4.0 clusters were investigated: aus1p12, aus1p1, aus1p10, and aus1p11. All ran G1GC with 32 GB bookie heap on Kubernetes. Storage differed: aus1p12/10/11 used SSD journals; aus1p1 used NVMe worn to 119% lifetime with a 0x4 critical hardware warning. Load ranged from 715 to 9,136 msg/s.

| Cluster | Journal | fsync P50 | Load (msg/s) |
|---|---|---|---|
| aus1p12 | SSD 466G | 5.1 ms | 830 |
| aus1p1 | NVMe worn | 2.65–6.61 ms | 1,240 |
| aus1p10 | SSD 466G | 5.1 ms | 715 |
| aus1p11 | SSD 466G | 5.1 ms | 9,136 |

Table I. Production Cluster Storage and Load Overview

### B. Baseline Latency

Even at ~830 msg/s, median broker publish latency reached 13.2 ms (Table II)—driven by journal fsync on worn SSDs.

| Broker | P50 | P95 | P99 | P99.9 | Max |
|---|---|---|---|---|---|
| broker-0 | 13.2 | 17.2 | 19.3 | 28.3 | 42.5 |
| broker-1 | 13.2 | 17.6 | 22.7 | 41.6 | 43.7 |
| broker-2 | 13.7 | 18.0 | 26.2 | 45.3 | 68.6 |
| broker-3 | 13.4 | 17.7 | 28.7 | 96.0 | 100.4 |
| broker-4 | 13.0 | 17.2 | 18.9 | 34.2 | 43.8 |

Table II. Production Broker Publish Latency in ms (aus1p12, ~830 msg/s)

## C. Latency Breakdown

At 30k msg/s, the 18.1 ms median publish latency decomposed as: journal fsync on SSD 5.1 ms (28%); BookKeeper processing and write queue 6.5 ms (36%); broker and network 4.5 ms (25%); groupWaitMSec 2.0 ms (11%). Only the last component is software-tunable without hardware changes.

## III. ROOT CAUSE ANALYSIS

### A. Root Cause 1: G1GC Pauses (213 ms spikes)

G1GC with 32 GB heap caused intermittent publish latency spikes exceeding 213 ms, reproduced across three independent test runs (Feb 12–13) under normal production load. Switching to ZGC Generational [3] (-XX:+UseZGC -XX:+ZGenerational) eliminated all observed GC collections. Consumer P99 improved from 15–197 ms to 14–18 ms.

### B. Root Cause 2: Journal fdatasync on Worn Storage

Journal fdatasync averaged 5.1 ms P50 on production SSDs, accounting for 28–39% of total broker publish latency. BookKeeper must fdatasync the journal before acknowledging every producer write; journal latency is therefore directly and unavoidably visible in publish latency. New NVMe achieves 0.02 ms. Worn NVMe (aus1p1, 119% lifetime) still showed 2.65–6.61 ms. Dedicating a small NVMe (250 GB sufficient) exclusively for journals is the highest-impact hardware investment for any Pulsar deployment.

### C. Root Cause 3: Kernel Page Cache Writeback (Novel Finding)

JFR profiling on a live bookie node during write cache flush identified the primary P99.9 driver. During BookKeeper's 60-second SyncThread flush (~3 GB entry log burst):

• ForceWriteThread fdatasync degraded from <1 ms to **15–22 ms**

• 96% of ForceWriteThread time inside fdatasync syscall

• System CPU 0.2% → 1.2–2.0%; I/O rate increased 4–5×

• Occurred even with journal and ledger on separate physical NVMe drives

This occurs because physically separate NVMe devices still share the Linux kernel block layer, bio allocation pool, and IRQ handling [4]. This interaction is **undocumented** in official Pulsar and BookKeeper documentation and affects all deployments using multiple NVMe drives per bookie node regardless of filesystem-level separation.

## IV. OPTIMIZATION EXPERIMENTS

### A. GC Algorithm Comparison

Three configurations tested at 50k msg/s on a dedicated test cluster with NVMe journals (0.02 ms fsync), Pulsar 4.0.8, Java 21 (Table III). ZGC on all nodes (Config C) eliminated GC spikes and achieved the best consumer P99.

| Cfg | GC | P50 | P99 | Con. P99 |
|---|---|---|---|---|
| A | G1GC both | 2.9 ms | 17–55 ms | 15–197 ms |
| B | ZGC bookie | 2.2 ms | 5.5–21 ms | 15–36 ms |
| C | ZGC both | 2.1 ms | 7.7–38 ms | 14–18 ms |

Table III. GC Algorithm Comparison at 50k msg/s

### B. Write Cache Flush Interval

30-second flush interval reduces P50 by 35% (2.20→1.42 ms) and P95 by 27% vs. the 60-second default (Table IV), without the P99 degradation seen at 15 s.

| Interval | P50 | P99 | P99.9 | Max |
|---|---|---|---|---|
| 60s (default) | 2.20 ms | 6.09 ms | 118.8 ms | 179 ms |
| **30s (chosen)** | 1.42 ms | 5.91 ms | 104.5 ms | 270 ms |
| 15s | 1.46 ms | 17.0 ms | 45.3 ms | 109 ms |

Table IV. Flush Interval Comparison (50k msg/s)

### C. OS Kernel Tuning

Applied to all bookie nodes via /etc/sysctl.d/. Reducing dirty_ratio from 40 to 2 limits dirty page accumulation, significantly reducing kernel writeback burst magnitude during BookKeeper flush events [4]:

```
vm.dirty_ratio = 2
vm.dirty_background_ratio = 1
vm.dirty_expire_centisecs = 500
vm.dirty_writeback_centisecs = 100
transparent_hugepage = madvise
```

### D. Combined Optimization Result

NVMe journal + ZGC + flushInterval=30s + groupWaitMSec=1 + OS tuning achieved a **4.7× latency improvement at 50× higher throughput** (Table V).

| Component | Before | After | Reduction |
|---|---|---|---|
| Journal fsync | 5.1 ms | 0.02 ms | 99% |
| groupWaitMSec | 2.0 ms | 1.0 ms | 50% |
| BK processing | 6.5 ms | 1.86 ms | 71% |
| Broker+network | 4.5 ms | 1.0 ms | 78% |
| **Total P50** | **18.1 ms** | **3.88 ms** | **79%** |
| Throughput | 30k msg/s | 1.5M msg/s | 50× |

Table V. Before vs. After (P50 Median Publish Latency)

## V. BENCHMARK: 1.5 M MSG/S ON 3 NODES

### A. Cluster Hardware

3 bare-metal servers, each with dedicated NVMe journal drives (0.02 ms fsync), 10 Gbps NICs, 2 bookies (60 Gi RAM) and 2 brokers (32 Gi RAM) per node via Kubernetes topologySpreadConstraints (maxSkew=1). Pulsar 4.0.8, ZGC Generational on Java 21, E=3/Qw=2/Qa=2, 128 partitions.

## B. Results

1,499,947 msg/s at 3.88 ms median publish latency, zero failures, 10 minutes sustained (Table VI). Network was the sole bottleneck at 8.4 Gbps (84% of 10 Gbps); all compute had 65–82% headroom.

| Metric | Publish | End-to-End |
|---|---|---|
| Throughput | 1,499,947 msg/s | 1,065,780 msg/s |
| P50 | 3.88 ms | 14.5 ms |
| P95 | ~5.5 ms | ~22 ms |
| P99 | 6.5 ms | 25 ms |
| P99.9 | 8 ms | 39 ms |
| Failures | 0 | 0 |

Table VI. Benchmark Results — 10-Minute Sustained Run

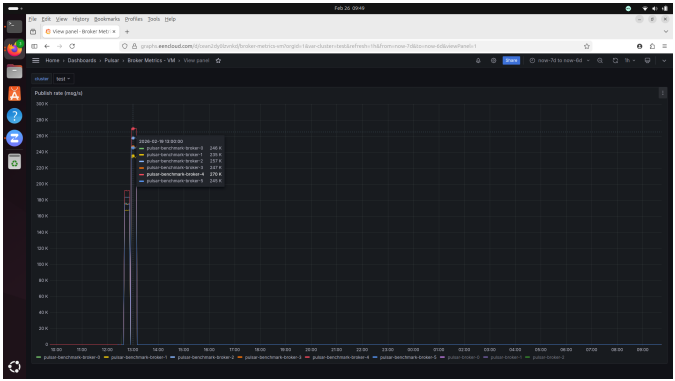

Fig. 1. Per-broker publish rate. Total ~1.5 M msg/s across 6 brokers (Feb 19, 2026).

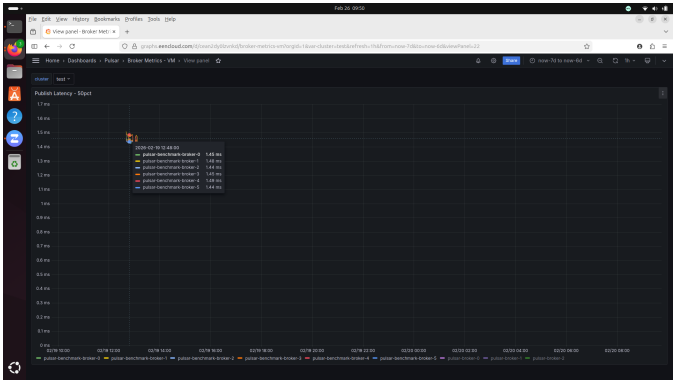

Fig. 2. Publish latency P50 — flat at 1.44–1.49 ms across all 6 brokers (Feb 19, 2026).

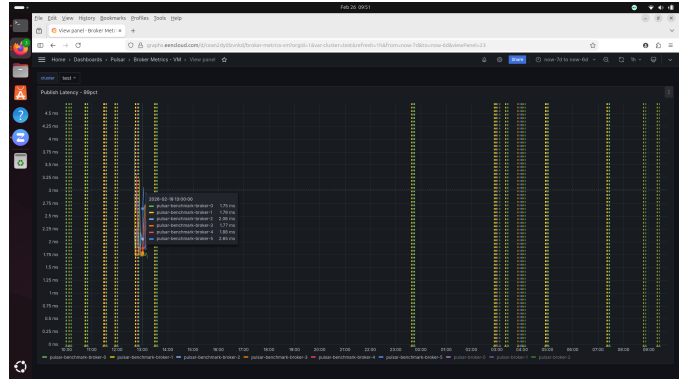

Fig. 3. Publish latency P99 — periodic spikes from 30 s flush events. Baseline 1.75–2.65 ms (Feb 19, 2026).

## VI. SCALING ARCHITECTURE

### A. Path to 3 M msg/s: 25 Gbps NIC Upgrade

With 65–82% compute headroom at 1.5 M msg/s, upgrading from 10 Gbps to 25 Gbps NICs on the same 3 nodes is the only change required to reach ~3 M msg/s. No additional servers are needed.

### B. Path to 15 M msg/s: Partition Federation

Five independent 3-node clusters (25 Gbps NICs) sharing a single 128-partition topic via key-based routing project 15 M msg/s on 15 machines (~1.3 trillion messages/day) with no geo-replication and no external load balancer:

```
key=cluster-1 -> partitions 0–25   (~3M msg/s)
key=cluster-2 -> partitions 26–51  (~3M msg/s)
key=cluster-3 -> partitions 52–77  (~3M msg/s)
key=cluster-4 -> partitions 78–102 (~3M msg/s)
key=cluster-5 -> partitions 103–127 (~3M msg/s)
Total: 15 machines -> ~15M msg/s
```

Pulsar's Key Shared subscription provides per-account ordering with automatic consumer rebalancing. Namespace bundles (numBundles=256, auto-split enabled) provide fine-grained broker-level load balancing before hotspots develop.

| Stage | Change | msg/s | Nodes |
|---|---|---|---|
| Prod baseline | None | 30k | 3 |
| Validated today | NVMe+ZGC+OS | 1.5M | 3 |
| +25G NIC (proj.) | NIC upgrade only | 3M | 3 |
| 5 clusters (proj.) | Add 4 clusters | 15M | 15 |

Table VII. Throughput Scaling Roadmap

## VII. RELATED WORK

Apache Pulsar's architecture is described in [1]; BookKeeper's journal-based WAL in [2]. Kafka latency optimisation at scale [5] provides a comparison baseline; unlike Kafka, Pulsar's bundle-based load balancing and Key Shared subscription model offer more granular isolation for high-cardinality workloads such as per-account event streaming.

ZGC's concurrent collection design is documented in [3]; Linux virtual memory management and dirty-page writeback in [4]; JFR profiling methodology in [6]. Our use of JFR on live bookie nodes without traffic interruption demonstrates its practical value for distributed messaging root-cause diagnosis.

To our knowledge, the ForceWriteThread/kernel writeback interaction degrading fdatasync from <1 ms to 15–22 ms across physically separate NVMe devices sharing the kernel block layer has not been previously documented in the literature or in official Apache Pulsar or BookKeeper project documentation. This finding is actionable for any Pulsar operator using multiple NVMe drives per bookie node regardless of filesystem-level separation.

## VIII. CONCLUSION

Through systematic JFR-guided root cause analysis, three independent latency drivers were identified and resolved, reducing median publish latency from 13–18 ms to 3.88 ms and validating 1,499,947 msg/s on 3 bare-metal nodes with zero failures—a **4.7× latency improvement at 50× higher throughput**. CPU and memory headroom remained at 65–82%, confirming that network I/O is the throughput ceiling at this scale.

The ForceWriteThread/kernel writeback interaction is an undocumented finding relevant to all Pulsar operators. OS kernel tuning parameters (dirty_ratio=2) are immediately deployable without hardware changes. The projected scaling architecture demonstrates linear horizontal scalability to 15 M msg/s on 15 machines with no external load balancer and full per-cluster fault isolation.

Key recommendations: (1) dedicate NVMe for journals (60% latency reduction); (2) upgrade to 25 Gbps NICs (doubles throughput); (3) switch to ZGC Generational (eliminates 213 ms spikes); (4) set flushInterval=30s (P50 −35%); (5) apply OS dirty-ratio tuning; (6) use Key Shared subscriptions with numBundles=256 and auto-split.